\documentclass[dvips]{article}

\usepackage{icrc2011}

\title{First results of the Standalone Antenna Array of the CODALEMA Radio Detection Experiment}

\newcommand{\etal}{\MakeLowercase{\textit{et al. }}} 
\shorttitle{Bell{\'e}toile \etal CODALEMA results}

\authors{Arnaud Bell{\'e}toile$^{1}$ for the CODALEMA collaboration$^{1,2,3,4,5,6,7}$}
\afiliations{$^1$SUBATECH, Universit{\'e} de Nantes/Ecole des Mines de Nantes/IN2P3-CNRS, Nantes France.\\
$^2$LESIA, USN de Nan{\c c}ay, Observatoire de Paris-Meudon/INSU-CNRS, Meudon France.\\
$^3$LPSC, Universit{\'e} 
Joseph Fourier/INPG/IN2P3-CNRS, Grenoble France.\\
$^4$LAL, Universit{\'e} Paris-Sud/IN2P3-CNRS, Orsay France.\\
$^5$GSII, ESEO, Angers France.\\
$^6$LAOB, Universit{\'e} de Besan{\c c}on/INSU-CNRS, Besan{\c c}on France.\\
$^7$LPCE, Universit{\'e} d'Orl{\'e}ans/INSU-CNRS, Orl{\'e}ans France.}
\email{arnaud.belletoile@subatech.in2p3.fr}

\abstract{CODALEMA is one of the pioneer experiments dedicated to the recent field of cosmic ray radio detection. It is located at the radio observatory of Nan{\c c}ay (France). 
The detector setup combined until recently a ground particle detector and an array of active dipole antennas covering a total area of~$0.25~\mathrm{km}^2$. 
The experiment is now going through a major upgrade with the deployment around the existing apparatus of a Standalone Antenna Array, which consists of 60 standalone new generation radio-detection stations and which will cover an area of $1.5~\mathrm{km}^2$~(33 stations deployed over the spring of 2011 and 27 stations to be deployed in late 2011). 
This new setup is intended to tackle the remaining unknowns of extensive air shower radio detection so as to make this technique a reliable and mature tool for Ultra High Energy Cosmic Ray (UHECR) physics. 
The latest results from the original CODALEMA array are discussed together with the first results of the Standalone Antenna Array.}
\keywords{ Cosmic ray, radio detection, CODALEMA, electric field topology, standalone detector, selftrigger}

\begin{document}
\maketitle

\section{Introduction}
A major upgrade is undergoing to make of the CODALEMA experiment a second generation detector of radio emission associated with cosmic rays. The following work summarizes the latest results of the original CODALEMA array and describes the new Standalone Antenna Array. 

\section{The original CODALEMA array}

\subsection{The setup and event selection procedure}

The original experimental setup of the CODALEMA instrument is detailed in~\cite{geomag}. It consists of two arrays of detectors : an antenna array of 24~fat active dipoles and a particle detector array of 13~plastic scintillators. Each detector (either antenna or scintillator) is connected to a central acquisition room by cables. The Nan{\c c}ay Decameter Array (DAM) which setup and results are detailled in~\cite{alxIcrc} is also fully integrated into the acquisition process.

Data acquisition is triggered when signal is detected in coincidence on the 5~central particle detectors within a $600~\mathrm{ns}$ time window. The offline analysis is performed independently on both arrays to ensure minimum bias. On one hand, radio transients are searched in antenna signals in a $23-83~\mathrm{MHz}$ frequency bandwidth. For each detected transient, the maximum electric field value and its arrival time are extracted. If more than 3~radio transients are detected in one event, the arrival direction of the electric field is reconstructed assuming a planar wavefront.

On the other hand, the arrival direction of the extensive air shower (EAS) is independently estimated from the relative arrival time in each particle detector assuming also a planar particle front. An analytical NKG lateral distribution~\cite{NKG} is then adjusted on the measured particle densities in the shower frame. If the shower core position stands inside the particle detector array, the event is referred to as internal and the fitted NKG distribution is used to estimate the energy of the primary using the constant intensity cut (CIC) method~\cite{gaisser} with an energy resolution of the order of $\Delta E/E \sim 30\%$ at $1\times 10^{17}~\mathrm{eV}$. 

Finally, an air shower candidate is identified by comparing arrival times and arrival directions obtained on each array. A coincidence smaller than $100~\mathrm{ns}$ in time and $20^\mathrm{o}$ in arrival direction validates the candidate as a radio detected EAS~\cite{nim}.

\subsection{Main features of the radio signal}

\begin{figure}[!ht]
  \vspace{5mm}
  \centering
  \includegraphics[width=2.5in]{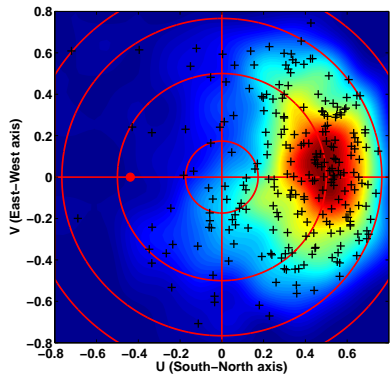}
  \caption{Black crosses show direction cosines on the ground plane of radio detected EAS using East-West polarised antennas. The geomagnetic vector is given by the red dot.  Concentric red circles indicate successively $10^{\mathrm{o}}$, $30^{\mathrm{o}}$, $50^{\mathrm{o}}$ and $70^{\mathrm{o}}$ zenith angle starting from the center of the figure. The superimposed smoothed density map highlights the North-South asymmetry.}
  \label{skymap}
 \end{figure}

One of the most appealing result of the CODALEMA experiment lies in the distribution of the arrival directions of radio detected showers. Whereas the cosmic ray flux is clearly isotropic around $10^{17}~\mathrm{eV}$, the radio detected events shown on figure~\ref{skymap} indicate a strong North-South asymmetry. Those events were detected using only the East-West polarisation of the electric field. The observation is well reproduced by assuming a first order emission process proportional to the Lorentz force, $\mathbf{F} = q.\mathbf{v} \times \mathbf{B}$ with $\bf{v}$ the arrival direction vector of the shower and $\bf{B}$ the geomagnetic vector.
The resulting vector is then projected on the East-West axis and convoluted with the scintillator array acceptance to reproduce a density map in good agreement with the one shown on figure~\ref{skymap}. Given the fact that the scintillator array energy threshold is relatively low, most of the radio detected EAS on CODALEMA belong to the transition region where the radio detection efficiency is below $100~\%$. Thus, for a given energy close to the threshold value, only events coming from the most favorable direction are seen by the antenna array and the asymmetry from figure~\ref{skymap} is understood as a threshold effect. Regarding the theoretical understanding of the emission process, this behavior confirms a dominant contribution of the geomagnetic effect~\cite{geomag}.

\subsection{The electric field topology}

\begin{figure}[!ht]
  \vspace{5mm}
  \centering
  \includegraphics[width=2.5in]{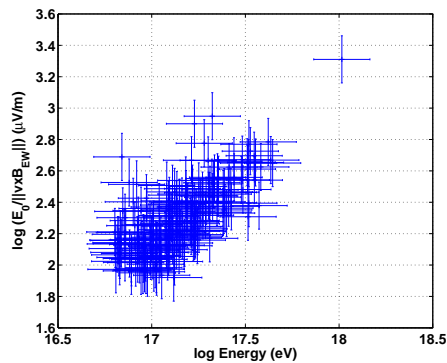}
  \caption{The normalised $E_0$ parameter for each fitted event is plotted as a function of the energy of the primary estimated with the particle detector array. The relative uncertainty of $30~\%$ for the radio estimator assumes a radio systematic error on the energy similar to the one of the particle detector.}
  \label{E0VsEcic}
 \end{figure}

On an event by event basis, we have shown in~\cite{astropart} that the measured electric field projected in the shower frame can be described by a profile of the form :

\begin{equation}
 E_i = E_0 \times e^{(-b_i/b_0)}
\label{profile}
\end{equation}

with $E_i$ the electric field measured by antenna $i$ at distance $b_i$ from the shower axis. When fitted to the data, this profile provides 4~parameters~: $E_0$, $b_0$ from above and $(X_0,Y_0)$ which correspond to the shower core position as seen by the antenna array.
A radio estimator of the energy of the primary is inferred from the fitted parameter $E_0$ normalised by a coefficient arising from the EAS arrival direction with respect to the geomagnetic vector as discussed above. The correlation of this estimator with the energy obtained from the scintillator array is shown on figure~\ref{E0VsEcic}. A dependence between the 2 estimators can already be seen, but some systematics remain and have to be determined in order to clean the energy correlation. The exponential profile, though suitable for a large part of events, may be questioned at this stage.

\begin{figure}[!ht]
  \vspace{5mm}
  \centering
  \includegraphics[width=2.5in]{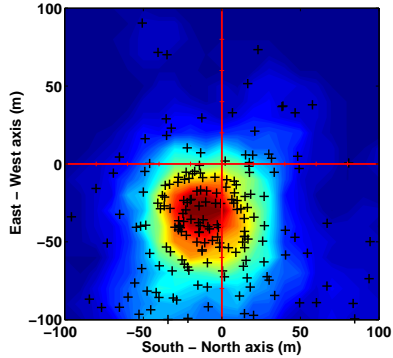}
  \caption{Difference between shower core positions obtained from radio signals and from the scintillator array. Each black cross is one EAS core position and a smoothed density map is superimposed. Instead, a systematical shift of the radio core of $\sim 20~\mathrm{m}$ toward the East is observed.}
  \label{dCore}
 \end{figure}

An additional interesting result lies in the reconstructed shower core position when fitting the electric field profile. As shown on figure~\ref{dCore}, the radio shower core positions, when compared to those obtained from the scintillator array, are, on average, sharply shifted toward the East direction. This difference was first mentioned in~\cite{alxIcrc}. Potential experimental biases originating from the geometry of the radio array or a trigger effect have been tested since then but failed to reproduce this core position shift. Thus a physical origin of this feature is to be considered. 

A possible interpretation of this shift as a signature of the charge excess contribution in the emission process is discussed in more details in this issue~\cite{vincent}. The observed difference is quantitatively compatible with the theoretical model discussed therein. Under such an hypothesis, a correlation is expected between this parameter and the incident angle of the shower compared to the geomagnetic vector. A marked variation of this shift is expected for showers arriving along, or closely to, the geomagnetic vector. The lack of events from this part of the sky does not allow to state upon this correlation with a high level of confidence yet. 

\section{The Standalone Antenna Array}

The original CODALEMA setup is currently being upgraded with an additional array of 60~standalone radio detectors. This new Standalone Antenna Array will cover an area of $1.5~\mathrm{km}^2$ that will allow us to focus on a higher energy range of $10^{17}$~to~$10^{18}~\mathrm{eV}$ where the radio detection reaches $100~\%$ of relative efficiency. 

\subsection{The elementary detector}

\begin{figure}[!ht]
  \vspace{5mm}
  \centering
  \includegraphics[width=2.5in]{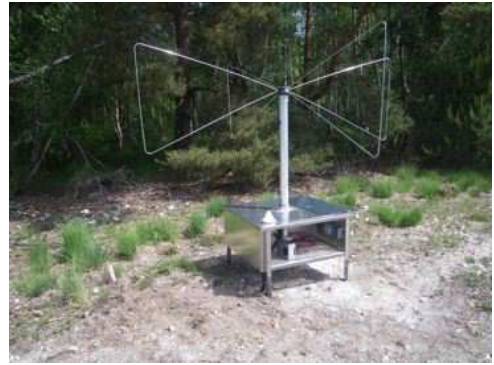}
  \caption{Picture of a standalone detector being deployed on the field.}
  \label{detector}
 \end{figure}

The key point of this new array lies in its elementary detector. The standalone station developed by the CODALEMA collaboration takes profit of the knowledge acquired on a first prototype that successfully achieved the first self triggered radio detection at the Pierre Auger Observatory~\cite{arena}. It provides data triggering, signal digitisation and processing, dating and communication with the outside world all by itself. One of those station is shown on figure~\ref{detector}. The new butterfly antenna~\cite{butterfly} associated to this station provides both East-West and North-South polarisation measurements with optimised sensitivity. 

Some of those detectors were validated on the CODALEMA array during a first phase of test. Another set of those stations was deployed at the Pierre Auger Observatory during spring 2010. The successful results of this array is presented in this issue~\cite{benoit}.

\subsection{The transient radio background}

\begin{figure}[!ht]
  \vspace{5mm}
  \centering
  \includegraphics[width=2.5in]{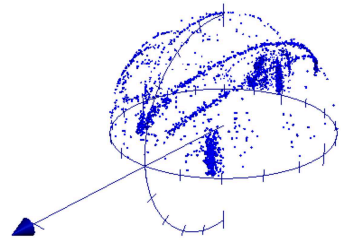}
  \caption{Arrival directions of radio transients detected with the Standalone Antenna Array. The arrow indicates the North direction. Each blue dot it a reconstructed event.}
  \label{reconstRadio}
 \end{figure}

One of the main issue a self-triggered radio array has to deal with is the transient radio background (or radio frequency interferences, RFI). During normal radio quiet periods, most of the RFI are seen in coincidence by several stations and can then be triangulated. A set of reconstructed RFI arrival directions acquired during 2 days of data taking is shown figure~\ref{reconstRadio}. One can easily distinguish several type of sources such as airplane tracks passing over the array. Those are of a particular interest as the knowledge of the position of a given airplane permits fine calibration of our instrument.

 Most of other RFI sources originate from the direct vicinity of the array. Those can be of various origin such as atmospheric effects or human made activities. Difficulties arise when RFI bring the detector trigger rate to its saturation limit. The observed CODALEMA transient radio background is highly variable from one day to the other and the array can be brought to saturation during bad periods. Several elaborated triggering approaches are currently under evaluation to improve the system RFI immunity.

\begin{table*}[]
\begin{center}
\begin{tabular}{c||c||c|c|c||c|c}
Date & Nb station & $\theta_{sc}(^{\mathrm{o}})$ & $\phi_{sc}(^{\mathrm{o}})$ &$\Delta \Omega(^{\mathrm{o}})$ & $\mathrm{\psi_{pred}}(^{\mathrm{o}})$ &  $\mathrm{\psi_{meas}}(^{\mathrm{o}})$\tabularnewline
\hline
2011/06/17 & 1 & 32.2 & -55.6 & - & 58 & 58\tabularnewline
2011/06/23 & 3 & 29.6 & -6.0 & 2.9 & 87 & 86-87-84\tabularnewline
2011/06/26 & 1 & 42.0 & 20.0 & - & 77 & 75\tabularnewline
2011/06/27 & 1 & 27.1 & 104.7 & - & 36 & 36\tabularnewline
\end{tabular}
\caption{Radio detected EAS candidates. From left to right : date of occurrence, number of radio stations in coincidence, $\theta_{sc}$ and $\phi_{sc}$ from the particle detector array, angular difference between radio and particle detector array, predicted $\psi$ using ($\theta_{sc}$,~$\phi_{sc}$) and measured $\psi$ for each radio station.}
\label{candidates}
\end{center}
\end{table*}

\subsection{The EAS radio detection}

The Standalone Antenna Array have been running into science mode in June 2011. During this period, 4 radio detected EAS candidates were found. Those EAS candidates, listed in table~\ref{candidates}, were identified by searching time coincidences between individual radio stations and the CODALEMA particle detector array within~$5~\mathrm{\mu s}$. 

For the event with 3 radio stations in coincidence, a firm discrimination between EAS radio signals and fortuitous coincidences is achieved in the exact same way it is done with the original CODALEMA array by comparing arrival directions obtained from radio signals to those extracted from the particle detector array. The resulting angular difference given in table~\ref{candidates} shows a good agreement between the 2 arrays and the candidate can be validated as a radio detected EAS.

A confirmation of EAS radio detection can also be brought for candidates with only one radio station. This is done by predicting the orientation of the electric field vector using the knowledge of its arrival direction. Indeed, it was shown in~\cite{geomag} that, for showers arriving away from the direction of $\mathbf{B}$, the electric field is at first order proportional to the Lorentz force vector. Thus, for a given arrival direction of the shower, the orientation of the electric field vector is predicted to be along $\mathbf{v} \times \mathbf{B}$ and the projection of this vector on the ground plane should have the same direction that the measured electric field. We define $\psi= \arctan (E_{EW}/E_{NS})$ the angle of this projected field with respect to the North direction. The table~\ref{candidates} shows a very nice agreement between predicted ($\psi_{pred}$) and measured ($\psi_{meas}$) angles for the 4 EAS events.  

The above picture is obviously a simplifying one and some systematics still remain to be quantified. The relative contribution of each emission mechanism, for instance, which is subject to variation depending on the arrival direction of the shower or atmospheric effects are to be considered precisely. 

Nevertheless, this identification criterion, though preliminary, opens the way of self sufficiency to the radio technique as a strong constrain on the EAS electric field can be inferred from the knowledge of its arrival direction only, regardless of any particle detector information. The new Standalone Antenna Array will bring the needed statistic for completeness.

\section{Conclusion}

Results of the original CODALEMA radio detection experiment have been reviewed. A shift of the shower core positions reconstructed with radio signals is observed when compared to those obtained from the scintillator array. This might be interpreted as a signature of the excess charge contribution to the electric field emission. 

The new Standalone Antenna Array is currently being completed. This new array provides refined measurements as well as increased statistics at higher energies. It already succeeded in EAS self triggered radio detection within a rather short time of operation.

On a physical point of view, a conclusion has to be reached on the ability to trace the radio measurements back to the cosmic ray composition. The energy estimator should be refined by investigating systematical effects and/or the exponential profile suitability. 

The EAS emission process picture becomes clearer as both, theoretical models and experiments, have known substantial progresses over the last few years. It makes no doubt that the dynamic community of radio detection will deliver soon an efficient tool for UHECR physics.


\clearpage

\end{document}